\documentclass[letterpaper,10pt,conference]{ieeeconf}

\usepackage{cite}

\def\BibTeX{{\rm B\kern-.05em{\sc i\kern-.025em b}\kern-.08em
    T\kern-.1667em\lower.7ex\hbox{E}\kern-.125emX}}
\usepackage{amsmath,amssymb,amsfonts}
\usepackage{mathrsfs}
\usepackage[ruled,vlined]{algorithm2e}
\usepackage{subcaption}
\usepackage{hyperref}
\usepackage[dvipsnames,svgnames,x11names]{xcolor}

\usepackage{tikz}
\usetikzlibrary{automata,positioning}
\usepackage{algorithmic}
\usepackage{graphicx}
\usepackage{subcaption}
\usepackage{textcomp}
\usepackage{acronym}
\usepackage{pgfgantt}
\usepackage[utf8]{inputenc}
\usepackage{amsmath}
\usepackage{amssymb}
\usepackage{bm}
\usepackage[ruled,vlined]{algorithm2e}
\usepackage[final]{pdfpages}

\usepackage{amsthm}
\usepackage{dirtytalk}
\usepackage{mathtools}
\usepackage{tikz}
\usetikzlibrary{arrows,automata}
\usepackage{romannum}
\usepackage{wrapfig}
\usepackage{todonotes}
\usepackage{siunitx}
\usepackage{lineno}
\usepackage{bbm}
\usepackage{subcaption}
\usepackage{tikz}
\usepackage{float}
\usepackage{graphicx}
\usepackage{color}
\usepackage{url}
\usepackage{multirow}
\usepackage{comment}
\usepackage{paralist}
\usepackage{scrextend}
\usepackage{amsfonts}

\newcommand{\ie}{{\it i.e.}}

\newcommand{\reals}{\mathbb{R}}

\newcommand{\norm}[1]{\lVert #1 \rVert}

\newcommand{\indicator}{\mathbf{1}}

\newcommand{\dist}[1]{\mbox{Dist}(#1)}

\acrodef{pa}[PA]{Probabilistic Automaton}
\acrodef{dfa}[DFA]{Deterministic Finite Automaton}
 \acrodef{mdp}[MDP]{Markov Decision Process} 
\acrodef{ltl}[LTL]{Linear Temporal Logic}
\acrodef{ltl3}[LTL$_3$]{ 3-valued LTL}
\acrodef{adp}[ADP]{Approximate Dynamic Programming}
\acrodef{tadp}[TADP]{Topological Approximate Dynamic Programming}
\acrodef{scltl}[sc-LTL]{syntactically co-safe LTL}
\acrodef{vi}[VI]{Value Iteration}
\acrodef{kl}[KL]{Kullback–Leibler}

\acrodef{pag}[PAG]{Probabilistic Attack Graph}
\acrodef{ids}[IDS]{Intrusion Detection System}

\newcommand{\sink}{\mathsf{sink}}
\newcommand{\regret}{\mathcal{R}}

\newcommand{\ids}{\mathsf{IDS}}

\acrodef{mtd}[MTD]{Moving Target Defense}
\theoremstyle{definition}
\newtheorem{definition}{Definition}[section]

\newtheorem{assumption}{Assumption}[section]
\newtheorem{remark}{Remark}
\newtheorem{problem}{Problem}

\newtheorem{example}{Example}[section]

\usepackage{xcolor}
\usepackage{booktabs}

\usepackage{todonotes}
\IEEEoverridecommandlockouts
\begin{document}

 \title{A Receding-Horizon MDP Approach for Performance Evaluation of Moving Target Defense in Networks}
\author{Zhentian Qian, Jie Fu,  and Quanyan Zhu %
\thanks{Z. Qian, J. Fu are with the Robotics Engineering Program and Dept. of Electrical and Computer Engineering, Worcester Polytechnic Institute, Worcester, MA 01609 USA.
{\tt\small \{zqian,jfu2\}@wpi.edu}}
\thanks{Q. Zhu is with  Dept. of Electrical and Computer Engineering, New York University, USA 
{\tt\small \{quanyan.zhu\}@nyu.edu}}
}

\thispagestyle{empty}
\pagestyle{empty}
\maketitle

\begin{abstract}
In this paper, we study the problem of assessing the effectiveness of a proactive defense-by-detection policy with a network-based moving target defense. We model the network system using a probabilistic attack graph--a graphical security model. Given a network system with a proactive defense strategy, an intelligent attacker needs to perform reconnaissance repeatedly to learn about the locations of intrusion detection systems and re-plan optimally to reach the target while avoiding detection.  To compute the attacker's strategy for security evaluation, we develop a receding-horizon planning algorithm using a risk-sensitive Markov decision process with a time-varying reward function. Finally, we implement both defense and attack strategies in a synthetic network and analyze how the frequency of network randomization and the number of detection systems can influence the success rate of the attacker. This study provides insights for designing proactive defense strategies against online and multi-stage attacks by a resourceful attacker.
\end{abstract}






\section{Introduction}
Cyber networks in industrial control systems are often targeted by malicious and resourceful attackers. An attacker can identify system vulnerabilities through reconnaissance and compromise the security of a network through calculated, multi-stage attacks. To counter the attacks, a network system can employ a mix of cybersecurity mechanisms, from traditional firewalls and intrusion detection to moving target defense \cite{senguptaSurveyMovingTarget2019} and cyberdeception \cite{jajodiaCyberDeceptionBuilding2016} with honeypots  \cite{DeployingHoneypotsHoneyd}. However, it is difficult to measure the effectiveness of dynamic defense techniques. The lack of understanding their security gains hinders the practical deployment of advanced dynamic defenses. 

Formal graphical security models, such as attack graphs \cite{jhaTwoFormalAnalyses2002} and attack-defense trees \cite{kordyFoundationsAttackDefense2011}, have been developed  \cite{schneierbruceAttackTrees2007} to evaluate security properties of a cyber system. An attack graph captures multiple paths that an attacker can carry out by exploiting vulnerabilities to reach the attack goal. Recent works \cite{hongAssessingEffectivenessMoving2016,hongHARMsHierarchicalAttack2012} have investigated the security property of \ac{mtd} using probabilistic attack graphs, where probabilistic transitions are uncertainties created by network-based randomization. However, there has not been an analytical model for evaluating the effectiveness of \ac{mtd} for  detection.

To detect the presence of an attacker, network administrators often place \acp{ids} at several points in the network to monitor traffic to and from all devices on the network and detect suspicious activities. They are essential components of proactive defenses, where  the defender is not aware of the existence of the attacker  but deploys some pre-defined security protocols.  The question we aim to address is that, given a proactive defense strategy  and an attacker who performs a sequence of actions to reach the target, as in lateral movement attacks \cite{NetworkLateralMovementa}, 
how effective is a proactive defense strategy to detect the attacker before the attacker succeeds?

For \ac{ids}s at fixed locations, an attacker can learn their locations during reconnaissance and avoid them while carrying out an attack. An effective detection technique, called ``roaming \ac{ids}s'', is used to randomize the location of \ac{ids}s in the network. For example, a flow-based \ac{ids} \cite{ids-sdn} allows network flow to  pass through and examined by \ac{ids} on a per-flow basis using software-defined networking.  Roaming decoys \cite{khattabRoamingHoneypotsMitigating2004} have also been used to mitigate Denial-of-Service attacks by shuffling the decoy locations in a network. This randomization creates uncertainty for the attacker and also increases his cost, as the attacker has to perform reconnaissance to determine the new \ac{ids} locations to avoid detection.

To understand how effective the defense strategy is, we need to understand how the attacker behaves given the uncertainty. To this end, we model the network with dynamic defense as a time-varying probabilistic attack graph, which can be modeled as a \ac{mdp} with a time-varying probabilistic transition function and a reward function. Then, we compute the attack strategy using risk-sensitive finite-horizon planning, and iteratively re-plan the attack strategy using a receding horizon framework. Given the computed attack strategy, we can evaluate the effectiveness of the detection and defense strategy by characterizing the relation among the probability of successful and stealthy attack, the number of \ac{ids}s, and the shuffling frequency of the \ac{ids}s.

Finally, the paper is structured as follows: In Section~\ref{sec:prelim}, we introduce preliminaries on attack graphs, roaming \ac{ids} defense strategy, and formulate the problem. In Section~\ref{sec:rhc_attack}, we design the receding-horizon attack planning in the time-varying network. In Section~\ref{sec:experiment}, we evaluate the performance of defense against the proposed online attacker planner. Section~\ref{sec:conclude} concludes the paper.

\section{Related Work}
In the context of moving target defenses, attack graph models \cite{Ehab19,Kim2015} and dynamic game models \cite{zhu2013game,Yunhan2020,chen2019dynamic,chen2019control,manshaei2013game} have been proposed to capture the strategic interactions between an attacker and a defender. In \cite{zhu2013game}, a multi-stage game has been proposed to model the kill chain of the adversary. In \cite{huang2020dynamic}, the authors have proposed a multi-stage game of incomplete information to model a long-term interaction of a proactive defender and a stealthy attacker. In recent work \cite{aslanyanQuantitativeVerificationSynthesis2016a,hansen2017quantitative,kordyQuantitativeAnalysisAttack2018},  attack-defense trees are developed to incorporate defender's countermeasure  \cite{kordy2010foundations} and capture the dependencies between actions and subgoals for both attacker and defender. These models are used to verify quantitative security properties expressed via temporal logic, based on the solutions of omega-regular games \cite{baierPrinciplesModelChecking2008,pitermanSynthesisReactiveDesigns2006,bloemGraphGamesReactive2018}. In \cite{huang2020strategic}, the authors have introduced online learning defense schemes that proactively interact with attackers to increase the attack cost and gather threat information. These approaches are applicable to synthesize reactive defenses: the defender is aware of the presence of the attacker and reacts to the attack actions in real time. In this work, we study proactive defense when the defender uses a fixed randomization strategy without knowing whether there is an attacker in the network.

For both reactive and proactive defense, one of the critical challenges in applying game theory to security is the performance evaluation of the attack behaviors. This work leverages a receding-horizon technique together with probabilistic attack graphs to assess the effectiveness of a class of cyber defenses that explicitly account for the attacker's uncertainties. The adversary model captures the key properties of the cyber kill chain \cite{rass2016gadapt,Tarun2015}, in which an attacker explores the network and its vulnerability, moves laterally in the network, and takes actions to achieve the attack goals, such as  data exfiltration, data destruction, or encryption for ransom. Performance evaluation is an essential first step toward the design of effective moving target defense. This work provides informative metrics that will be useful to address issues related to defense design, resource planning, and security investment.





\section{Preliminaries and Problem Formulation}
\label{sec:prelim}
 In this section, we present preliminaries on formal graphical security models, and then formulate the problem to evaluate the effectiveness of the proactive defense strategy.


\begin{definition}[Probabilistic attack graph]
A \ac{pag} is a probabilistic transition system $G = \langle S, A, P, s_0\rangle$
where $S$ is a set of network nodes, $A$ is a set of attack actions, and $P: S\times A\rightarrow \dist{S}$  is a probabilistic transition function--that is, $P(s'|s,a)$
 is the probability of the attacker reaching node $s'$  from a (compromised) node $s$ with an attack action $a$ (targeted at $s'$ only). The probability of failing to exploit a vulnerability results in a self-loop $P(s|s,a) = 1-  P(s'|s,a)$. The state $s_0$ is the initial entry node for the attacker.
 \end{definition}
 The reader can think of the \ac{pag} as an \ac{mdp}, in which the set of actions are the attacker's exploitation actions.  The probability  of an attacker successfully exploiting a vulnerability 
 can be estimated based on the Common Vulnerability Scoring System (CVSS)  \cite{CommonVulnerabilityScoring}, as used in
\cite{frigaultMeasuringNetworkSecurity2008,munoz2017exact}.
 
 
Using network-based \ac{mtd} techniques, we can randomize the software/hardware or the topology of the network. We consider a case of \ac{ids} randomization techniques where the locations of \ac{ids}s can be sampled from the set of nodes of the network.  For example, if $s\in S$ is sampled, then all flows into node $s$ will be examined by an \ac{ids} and we say that node $s$ is equipped with an \ac{ids}. 

For simplicity, we assume that when the attacker sends a package to exploit the vulnerability of a target node and the node is equipped with an \ac{ids}, the attacker will be detected and blocked from the network. 


We aim to evaluate the security level of the system for a proactive defense strategy, defined as follows.
\begin{definition}
A \emph{periodic} defender strategy $\delta(t+T_r) =\delta(t)$ that randomly selects $k$ out of a subset $\cal N \subseteq S$ of nodes in the network as  the \ac{ids} locations every $T_r$ steps.  
\end{definition}
\begin{assumption} 
The following assumptions are made for the attacker:
 \begin{itemize}
     \item The attacker knows the \ac{pag} but does not know the defender strategy $\delta$ and $T_r$.
     \item The attacker can exercise the network scan every step, before taking any attack action, to learn about the locations of \ac{ids}s at that moment.
     \item The defender's action of sampling \ac{ids}s  is taken concurrently with the attack actions.
 \end{itemize}
\end{assumption} 
It is noted that if the defender uses a Poisson distribution over the period $T_r$, even if the attacker learns the mean and variance, he cannot know exactly when the \ac{ids}s have been shuffled. Thus, the assumption that the attacker does not know $T_r$ is not necessary. 
 If we assume that the attacker knows the defender's strategy, then the attacker's planning problem reduces to a standard \ac{mdp}  whose solution provides the worst case analysis of the network defense. In this work, we are interested in studying how the attacker's lack of information can lead to a less conservative assessment of defense strategy.

\begin{definition}[Reach-avoid attack objective]Given the \ac{pag} and let $s_f\in S$  be the target for the attacker, the objective of the attacker is to avoid detection until reaching $s_f$. 
\end{definition}

\begin{definition}[Detection events]
An attacker can be detected if he attempts any action $a\in A$ at node $s$ to reach target $s'$, and $s'$ is equipped with an \ac{ids}.
\end{definition}
In other words, the attacker can be detected by exploiting a node equipped with an \ac{ids}, no matter whether the attack action is successful or not.

\begin{example}
We introduce an example to illustrate the concept.  Figure~\ref{fig:network-ex} depicts a small network with three hosts, equipped with SDN-enabled roaming \ac{ids}. At each time step, the \ac{ids} can be randomly assigned to a target host and monitor the flow.  Figure~\ref{fig:attack-noids} shows a transition in the \ac{pag} where the attacker has gained trust on host $1$, which is an FTP server. The FTP server consists of a vulnerability
which allows the attacker to obtain reverse shell (rsh) on
the system. By carrying out $\texttt{rsh}$ attack on host $1$, the attacker succeeds with probability $p$ to gain user access on host $1$, and with probability $1-p$ that his action fails. When the IDS is equipped with host 1, then the attacker's action $\texttt{rsh}$ will be detected, leading to the sink state--$\mbox{detected}$--in Fig.~\ref{fig:attack-ids}. 
 \begin{figure}[t]
    \centering
    \includegraphics[width=0.25\textwidth]{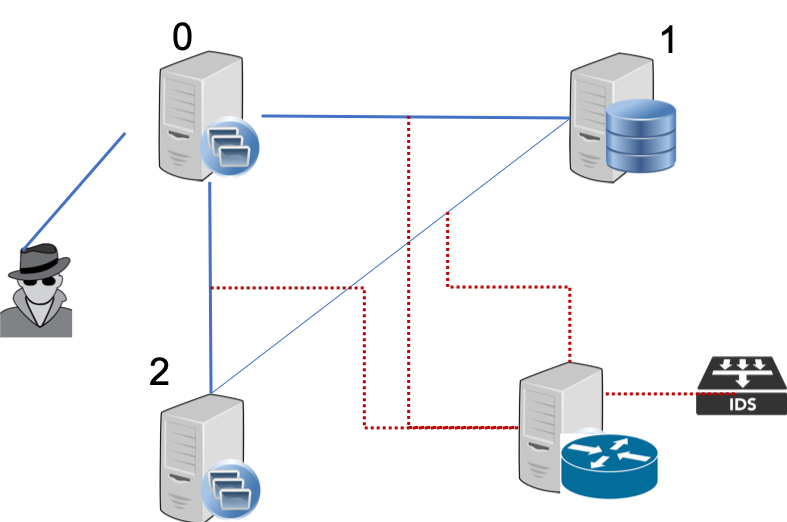}
    \caption{An example of a small network with roaming \ac{ids}.}
    \label{fig:network-ex}
\end{figure}
\begin{figure}[t]
    \centering
    \begin{subfigure}[b]{0.5\textwidth}
    \centering
    \includegraphics[width=0.85\textwidth]{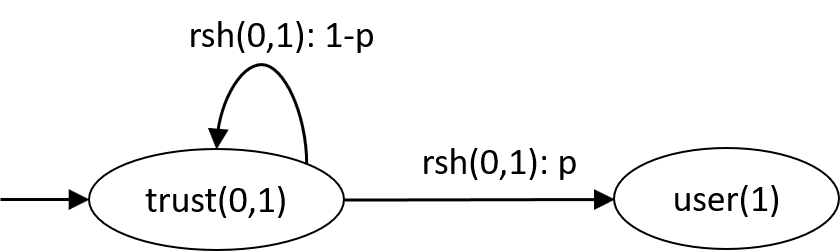}
    \caption{Attack graph with no \ac{ids}.}
    \label{fig:attack-noids}
    \end{subfigure}
    \begin{subfigure}[b]{0.5\textwidth}
    \centering
    \includegraphics[width=0.85\textwidth]{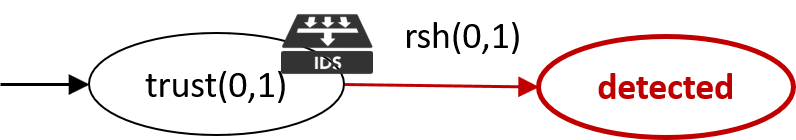}
    \caption{Attack graph with \ac{ids}.}
    \label{fig:attack-ids}
    \end{subfigure}
     \caption{A fragment of the \ac{pag}: (a)  Without \ac{ids}, the attacker carries out action to reach host 1 with some probability. (b) With \ac{ids}, the attack action is detected.}
    \label{fig:attackgraph-ex}
\end{figure}
\end{example}

\begin{problem}
Given a defense strategy $\delta$ and an initial state $s_0\in S$ of \ac{pag}, with what probability can the attacker achieve his attack objective? What is the best response of the attacker given the lack of knowledge in the defender's strategy?
\end{problem}

\section{Attacker's behavior modeling}
\label{sec:rhc_attack} To understand how the attacker plans given the non-stationary environment, we introduce an attack behavior model using online planning in \ac{mdp}s. In this section, we first introduce a preliminary on risk-sensitive, finite-horizon planning, and then present a receding horizon framework that iteratively solves finite-horizon problems in a time-varying \ac{mdp}.

\subsection{Preliminaries: Risk-sensitive planning in \ac{mdp}s}
  Given an \ac{mdp} $G= (S,A, P, 
 s_0)$, where $S, A, P$ are state, action spaces and transition function, respectively; $s_0$ is the initial state. We introduce an immediate reward function as:
\begin{equation}
    r_t:S \times A\rightarrow \reals^+, \; \forall \; t\in [t_0,t_0+ T-1], 
\end{equation}
where $T\ge  0$ is a constant for finite horizon length. The terminal reward $r_{t_0+T}:S\rightarrow \reals^+$ depends only upon the state $s \in S$.
The finite-horizon risk-sensitive optimal planning problem is described as follows: Given the \ac{mdp}, the immediate reward function $r_t, t\in [{t_0},t_0+ T-1]$ and the terminal reward function $r_{t_0+T}$, compute a policy  $\Pi ^{t_0} = (\pi_{t_0}, \pi_{t_0+ 1},\ldots, \pi_{t_0+ T-1})$ where $\pi_t :S\rightarrow \textrm{Dist}(A)$ maximizes the following objective:
\begin{multline}
    \label{eq:risk-sensitive-objective}
    {J}_{t_0}(\nu, \Pi^{t_0}) =\\ E^{\nu, \Pi^{t_0}} \big[\exp\big(\lambda \sum_{n={t_0}}^{t_0 + T-1} r_n(S_n,A_n)\\+r_{t_0+T}(S_{t_0+T})\big)\big],
\end{multline}
where $\lambda$ is a discounting factor; $\nu$ is the   distribution over states at $t={t_0}$, in our case it only resides on a single state $s_0$; the expectation $E^{\nu, \Pi^{t_0}} $ is computed from the Markov chain induced using policy $\Pi^{t_0}$; i.e., the state and action processes $\{S_t\}_{{t_0}\leq t \leq t_0+ T}$, $\{A_t\}_{{t_0}\leq t \leq t_0+ T-1}$.

As shown in \cite{kumar2015finite}, the risk-sensitive objective can be minimized using linear programming, with the primal and dual linear programs formulated as follows.

\medskip

\noindent\textbf{Primal Linear Program: }

\begin{align}
\label{eq:primalLP}
     \begin{split}
    & \min_{\big\{\{u_t(s)\}_{s \in S,{t_0}\leq t \leq t_0+ T -1}\big\}}\sum_{s \in S}\nu(s)u_{t_0}(s),  \\
\mbox{subject to:}&\\
     & u_{t_0+ T-1}(s) \geq b_{s,a}, \quad \forall s\in S, \forall a \in A,\\
& u_{t}(s)  - e^{r_t(s,a)}\sum_{s' \in S}P(s'|s,a)u_{t+1}(s')\geq 0,\\ 
& \forall \; s\in S, \forall a \in A, \forall t: {t_0}\leq t\leq t_0+T -2,
\end{split}
\end{align}
 where
\begin{equation}
b_{s,a}:=e^{r_{t_0+T-1}(s,a)}\sum_{s' \in S}P(s'|s,a)e^{r_{t_0+T}(s')}.
\end{equation}
The solution of the primal LP provides $\{ u_{t}(s) \mid s\in S, t_0\le t \le t_0+T-1\}$, where $u_t(s) = \max_{ \Pi^t} J_{t}(s, \Pi^t )$ (see \eqref{eq:risk-sensitive-objective}) with $\Pi^t = [\pi_t,\ldots, \pi_{t-t_0+T-1}]$.

\medskip
\noindent\textbf{Dual Linear Program: }
\begin{align}
\label{eq:dualLP}
     \begin{split}
    & \max_{y} \sum_{a \in A}\sum_{s \in S}b_{s,a}y(t_0+T-1,s,a)\\
 \mbox{subject to:}&\\
& \sum_{a\in A}y({t_0},s',a) = \nu(s'),\quad \forall s' \in S,\\
 & \sum_{a\in A}y(t ,s',a) = \\ &\sum_{a\in A}\sum_{s\in S}e^{r_{t-1}(s,a)}P(s'|s,a)y(t-1,s,a)\\
  &\forall t: t_0+1 \leq t \leq t_0+ T -1, \; \forall s' \in S.
  \end{split}
  \end{align}

Here, the decision variables $y$ are taken as $y=\{y(t,s,a)\mid {t_0}\leq t\leq  t_0+ T-1\}$. The solution to the dual LP would define the optimal policy of the risk sensitive MDP: For each  $t$ such that $t_0 \leq t \leq t_0+ T -1$, the nonstationary policy is 
\begin{equation}
\pi_{t}(s,a):=\frac{y(t,s,a)}{\sum_{a'}y(t,s,a')}, \forall\; s \in S \text{ and } \forall a \in A. 
\end{equation}

\subsection{Receding-horizon attack planning}

The receding-horizon model captures the lateral movement of the reconnaissance-exploitation-actions kill chain of an attacker. At each horizon, the attacker intends to map out the locations of \ac{ids}s in the network using reconnaissance techniques. Then, the attacker exploits the vulnerability to act and move to the next node. This process iterates until the attacker reaches his target.

At each step $t$, the attacker solves an \ac{mdp} with  set $S_{\ids,t} \subseteq S$ of nodes equipped with \ac{ids}s. We treat these nodes as obstacles which the attacker aims to avoid.   Given the \ac{mdp} $ ( S, A,P, s_t)$ with \ac{ids} placing at $S_{\ids,t} \subseteq S$ and the current state $s_t$, the reward function  is defined as follows:
\begin{equation}
\label{eq:reward}
    r_{t+k}(s,a)=0, \;\forall s\in S, \forall a\in A; \forall k\in [t,t+T-1];
\end{equation}
and 
\begin{equation}
\label{eq:term-reward}
r_{t+T}(s)  = \left\{\begin{array}{cc}
 1    &  \text{ if } s=s_f;\\
 0    &  \text{otherwise}.
\end{array}\right.
\end{equation}
In addition, let $\sink$ be an absorbing state with zero  reward. The transition function is revised as follows: For each $s\in S $, for each $a\in A$, if $P(s'|s,a) >0$  and $s'\in S_{\ids,t}$, then $P(\sink|s,a)=1$.  In other words, when the attacker exploits a vulnerability that has a positive probability to reach a node with \ac{ids}, then he will reach a sink state with probability one--that is, he is detected. 

\begin{remark}
It is noted that the detection occurs due to the concurrency of actions by the defender and an attacker. If the attacker always knows where the \ac{ids}s are in the next moment, then he can avoid these \ac{ids}s by either doing nothing or exploits vulnerabilities only on hosts that are not equipped with \ac{ids}s. However, randomization and concurrency together create the unknown effects when the attacker exploits. 
\end{remark}

\begin{remark}
 The length of the planning horizon $T$ is assumed to be fixed. However, in practice, it can depend on the dynamic tempo of the dynamic defense and attacker's computational resources. Future work will consider attackers with bounded rationality \cite{simon1990bounded}. In this paper, we examine one-time interaction, where the attacker does not have enough data to learn the defender's strategy. Adaptive attacker who can learn the defense strategy must collect data from multiple  interactions. 
\end{remark}

 This receding-horizon attack planner is described in  Alg.~\ref{alg: RHC}. It starts with $t=0$, the attacker scans the network and determines the location  $S_{ \ids,t}$ of \ac{ids}s. Then, the attacker generates the reward function $r_{t+k}$ and $r_{t+T}$ and solves the finite-horizon risk-sensitive \ac{mdp} and obtain the policy $\Pi^t$. The attacker then takes an action $a_t$ from the policy. This process iterates until either the attacker reaches the goal or becomes detected.
 
\begin{algorithm}
  \caption{The receding horizon attack planning algorithm}
  \label{alg: RHC}
    \KwIn{The \ac{pag} with initial state $s_0$ and target $s_f$. Finite planning horizon $T$ and total attack horizon $T_{\max}$.}
    \KwOut{$\pi_t$ at each time step $t = 0 \dots T_{max}$.}
  \begin{algorithmic}[1]
    \STATE (Initialization): $t = 0$.
    \WHILE{$t <T_{max}$}
        \STATE Netscan, obtain $S_{\ids,t}$;
        \STATE  Get rewards $r_{t+k}, r_{t+T}$ from 
        $S_{\ids,t} $ with \eqref{eq:reward} and \eqref{eq:term-reward}.
        \STATE Solve $\Pi^t = \{\pi_t,\pi_{t+1},\ldots, \pi_{t+T-1}\}$ with \eqref{eq:dualLP}.
        \STATE Take action $a_t \sim \pi_t(s_t)$ to reach $s'$. 
        \newline
        The \ac{ids}s replaced at $S_{\ids, {t+1}}$.\COMMENT{The network topology changes.}
        \IF{$s' \in S_{\ids,t+1}$}
        \STATE  Break. \COMMENT{Attacker is detected.}
        \ELSE
        \STATE With probability $p$, reach $s'$, $s_{t+1} \leftarrow s'$; \STATE With probability $1-p$, stay $s_t$,  $s_{t+1}\leftarrow s_t$.
        \ENDIF
        \IF{$s_{t+1} =s_f$}
        \STATE Break. \COMMENT{Attacker succeed.}
        \ELSE
                 \STATE $t\leftarrow t+1$; \COMMENT{Time increment.}
                 \ENDIF
    \ENDWHILE
  \end{algorithmic}
\end{algorithm}

Given that the attacker uses an online planner, the performance can be evaluated based on regret. To evaluate this regret, we need to solve the optimal policy of the attacker assuming the attacker knows exactly the sequence of locations for \ac{ids}s sampled over his planning horizon. This optimal policy can be obtained from the following \ac{mdp} as a stochastic shortest path problem, described below.

\begin{definition}
Given an \ac{mdp} $G =(S,A,P,s_0)$ and the attacker's goal state $s_f$, let $[S_{\ids,0}, S_{\ids,1},\ldots, S_{\ids, T_{\max}}]$ be a sequence of sampled subsets of nodes equipped with $\ids$s over the time horizon $[0,T_{\max}]$. 
A time-augmented \ac{mdp} $\tilde G = \langle S\times [0,1, \ldots, T_{\max}]\cup \{\sink \}, A, \tilde P, (s_0,0), \tilde r \rangle $ is defined as follows: 
$S\times [0,\ldots, T_{\max}]\cup \{\sink \}$ are the set of states, $A$ is the set of actions, $(s_0,0)$ is the initial state. The transition function is defined as: For each $t\in [0, T_{\max}-1]$,  each $a\in A$,  and each $s\in S$, there are four cases:
\begin{enumerate}
    \item If $s \neq s_f$,     $P(s'|s,a) >0$, $s' \notin S_{\ids, t+1}$ and $s'\ne s$, then let $ \tilde P((s',t+1)|(s,t),a)  = P(s'|s,a)$ and $\tilde P((s,t+1)|(s,t),a) = P(s|s,a)$. 
 \item If  $s=s_f$, let
$
\tilde P(\sink|(s,t), a) =1,
$ where $\sink$ is an absorbing state for any action $a \in A$.
\item If    $P(s'|s,a) >0$, $s' \in S_{\ids, t+1}$, and $s\ne s'$, then let $\tilde P(\sink|(s,t),a)  =1.$
\item If $t= T_{\max}$, 
$\tilde P(\sink |(s,T_{\max}), a) =1.$

\end{enumerate}

The reward function is defined by 
\begin{equation}
\tilde r(s,t)= \indicator(s\equiv s_f).
\end{equation}

\end{definition}
Let $\tilde \pi^\ast$ be the optimal solution of $\tilde G$ that maximizes the following objective:
\begin{equation}
J((s_0,0), \tilde \pi) = E^{(s_0, 0),\tilde \pi}\left[
\exp \left(\lambda \sum_{n=0}^h r((s,n), a_n)\right) 
\right]
\end{equation}
where $h$ is the first time when the policy-induced chain reaches the sink state.

Let $\Pi^0 = [\pi_0,\pi_1,\ldots \pi_h]$ be the sequence of policies performed by the attacker using the receding horizon planning with a finite horizon $h$.
Based on the solution of time-augmented \ac{mdp} $\tilde G$, we can compute the dynamic regret: 
\begin{equation}
\regret(\Pi^0) = \norm{J((s_0,0), \Pi^0) -J((s_0, 0), \tilde \pi^\ast)},
\label{eq:regret-formula}
\end{equation}
where $J((s_0,0), \Pi^0)$ is the evaluation of the policy in the time-augmented \ac{mdp} $\tilde G$, $J((s_0, 0), \tilde \pi^\ast)$ is the reward that can be obtained in the finite horizon by executing optimal policy $\tilde \pi^\ast$. The dynamic regret captures the performance difference  of policy $\Pi^0$ and optimal policy $\tilde \pi^\ast$. 
We will use dynamic regret to analyze the performance of the defense strategy. The proposed attack planning algorithm does not learn and predict the changes in the network. Thus, it does not minimize the dynamic regret.
In the future work, we will consider online attack learning-based  planning with regret minimization.

\section{Experiments}
\label{sec:experiment}
\subsection{Experimental setup}

We implement the proactive defense strategy in a synthetic network and the proposed attack planning algorithm to evaluate how effective the   defense strategy is. 
 All experiments in this section are performed on a computer equipped with an Intel R Core\textsuperscript{TM} i7-5700HQ and 8GB of RAM running a python 3.6 script on a 64-bit Ubuntu R 18.04 LTS.

\begin{figure}[htbp]
  \centering
  \includegraphics[width=.9\linewidth]{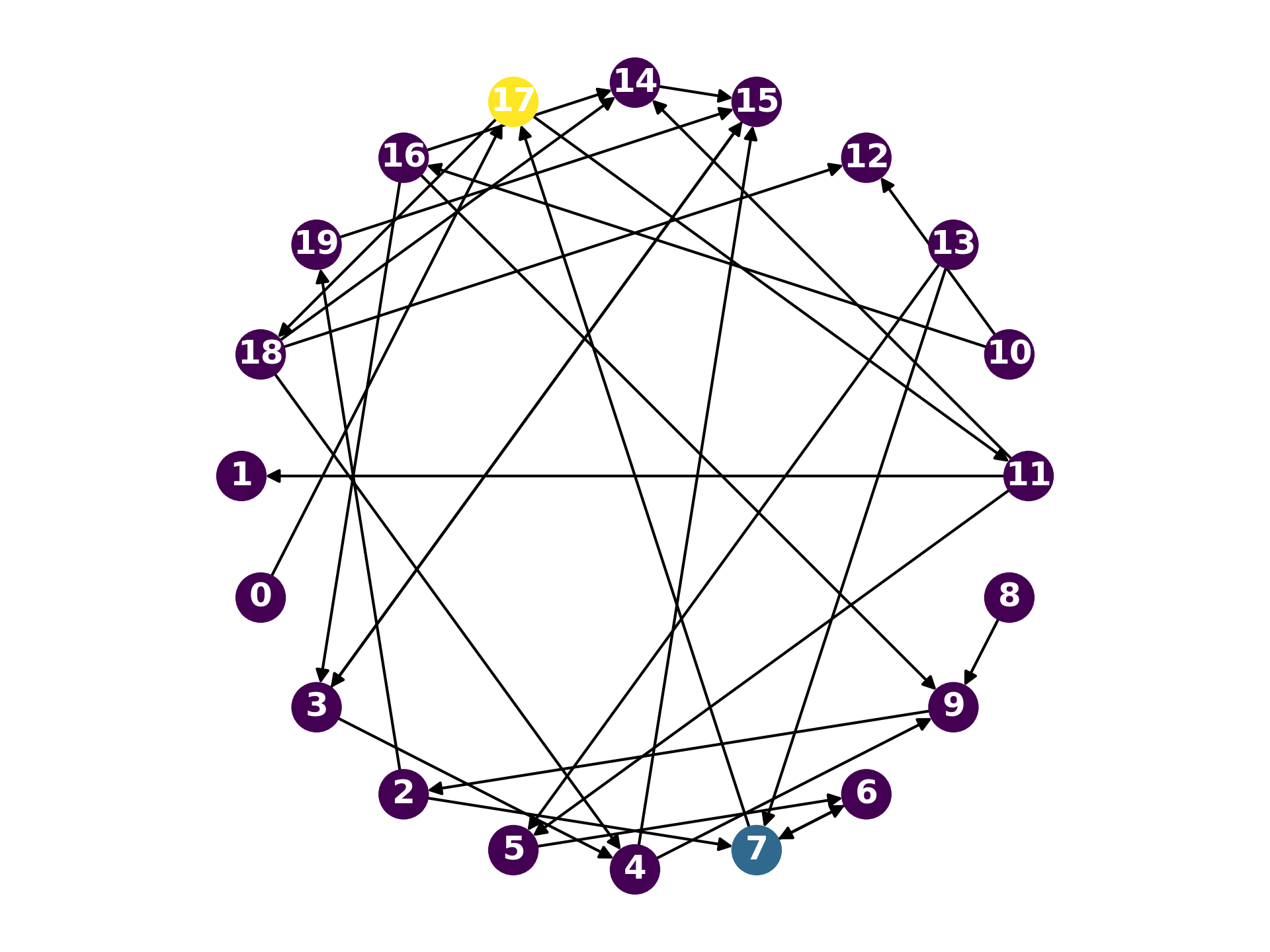}
  \caption{The layout of the probabilistic attack graph from a synthetic network.}
  \label{fig:network graph}
\end{figure}

The layout of \ac{pag} from the synthetic network is  shown in Fig. \ref{fig:network graph}.  The graph has twenty nodes. Note that the self-loops are omitted in the graph for clarity. The \ac{ids}s in the network are sampled using a random sampling process using a uniform distribution from subset $\mathcal{N}=\{0, 12, 2, 8, 1, 13, 15, 10, 9, 5\}$ at every $T_r$ steps, i.e., sampled at $\frac{1}{T_r}$ frequency. When $T_r$ approaches infinity (\ie, $0$ frequency), the locations of \ac{ids}s do not change. The attacker does not know $T_r$ and recomputes his policy every step. We assume that once the \ac{ids}s are selected, the attacker knows their new locations of \ac{ids}s after scanning the network. Thus, the analysis using this type of attacker provides a lower bound on the security level of the system, measured by the probability that the attacker can reach the target while avoiding \ac{ids}s. 

We conduct an experiment to investigate how the effectiveness of the roaming \ac{ids}s policy can be influenced by (1) the frequency in re-sampling and (2) the number of \ac{ids}s. In the experiment, the number of \ac{ids}s in the network ranges from one to five. The frequency of the sampling of the \ac{ids}s ranges from zero (i.e., the location of the \ac{ids}s never changes) to one (i.e., the locations of the \ac{ids}s change every time instant). Table \ref{tab:parameters} shows the parameters used in the attacker's receding horizon planning.

\begin{table}[htbp]
\centering
\caption{Experiment parameters}
\label{tab:parameters}
\begin{tabular}{ll}
\toprule
Parameters &  Values\\
\midrule
Finite horizon length $T$ & 19 \\
Maximum time length $T_{max}$ & 100 \\
Probability of successfully exploit a vulnerability $p$ & 0.9 \\
Attacker initial state $s_0$ & 17 \\
Attacker target state $s_f$ & 7 \\ 
Discounting   factor $\lambda$ in \eqref{eq:risk-sensitive-objective} & 1.0\\
\bottomrule
\end{tabular}
\end{table}

\subsection{The frequency of the re-sampling of the \ac{ids}s}

\begin{figure}[htbp]
  \centering
  \includegraphics[width=.9\linewidth]{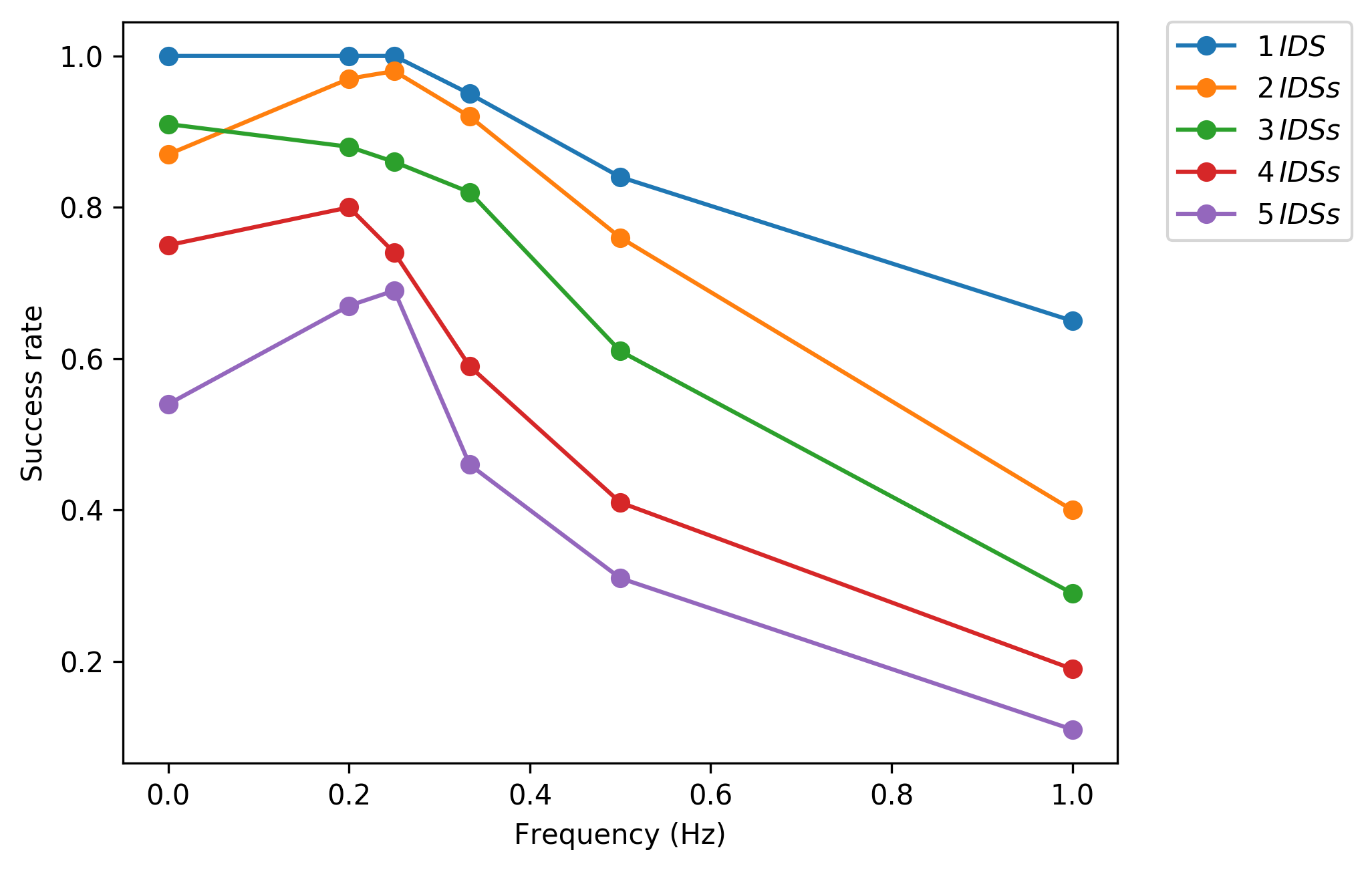}
  \caption{The effect of the frequency of the re-sampling of the \ac{ids}s on the success rate of the attacker.}
  \label{fig:frequency}
\end{figure}
The experiment results are shown in Fig. \ref{fig:frequency} and Fig.~\ref{fig:num_ids}. 
 From Fig. \ref{fig:frequency}, it is observed that the success rate of the attacker reaching the target decreases as the re-sampling frequency increases. The results are intuitive as a higher sampling frequency leads to a higher probability of an attacker reaching an IDS. However, the more frequent shuffle of \acp{ids} may incur overhead costs including traffic delay and disruption. It is also interesting to observe that the success rate of attack when re-sampling at $\frac{1}{5}$Hz is higher than that at a frequency of zero.  This is because re-sampling would sometimes free the attacker from a deadlock. For example, when the attacker is at state 0 and the \ac{ids} is at state 17, the best strategy for the attacker is to remain put. When the \ac{ids}s are being re-sampled every $T_r$ steps, the deadlock is lifted. However, the same observation may not be obtained if the \ac{ids}s are located at different nodes initially or the attacker starts with different initial nodes in the network. 
 The choice of sampling locations of \ac{ids}s  requires game-theoretic reasoning using, for example,  resource-allocation games \cite{korzhyk2010complexity}, and it will be analyzed in the future work.

\subsection{Number of \ac{ids}s}

\begin{figure}[htbp]
  \centering
  \includegraphics[width=.9\linewidth]{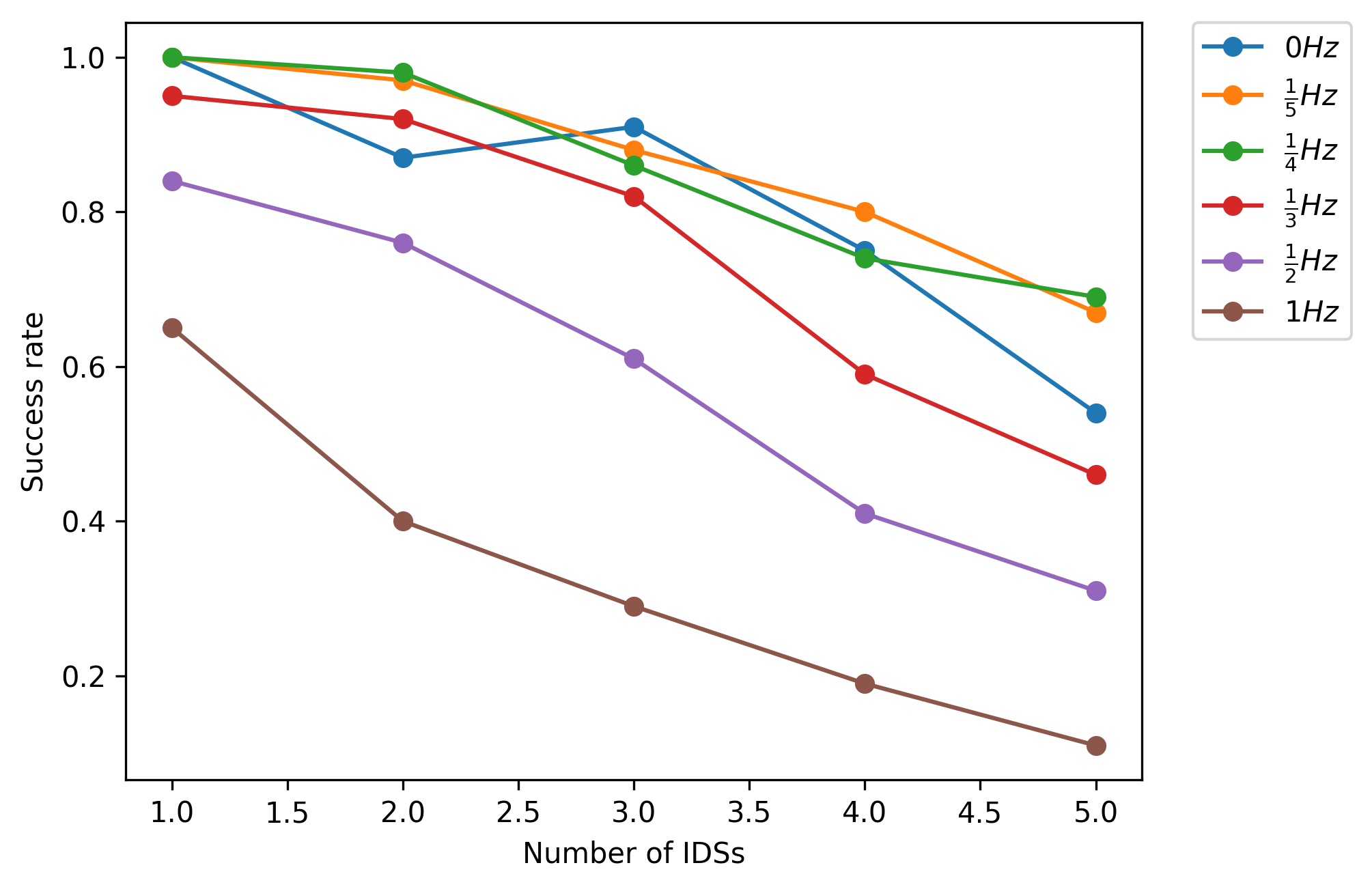}
  \caption{The effects of the number of \ac{ids}s on the success rate of the attacker.}
  \label{fig:num_ids}
\end{figure}

In Fig.~\ref{fig:num_ids}, we show the experiment results that describe how  the number of the \ac{ids}s in the network influences the effectiveness of the roaming \ac{ids} policy.  

From Fig. \ref{fig:num_ids}, it can be seen that the success rate of the attacker reaching the target decreases as the number of the \ac{ids}s in the graph increases. It suggests that a higher number of IDSs leads to a more effective roaming IDS policy. 

\subsection{The distance to the target}
\begin{figure}[htbp]
  \centering
  \includegraphics[width=.9\linewidth]{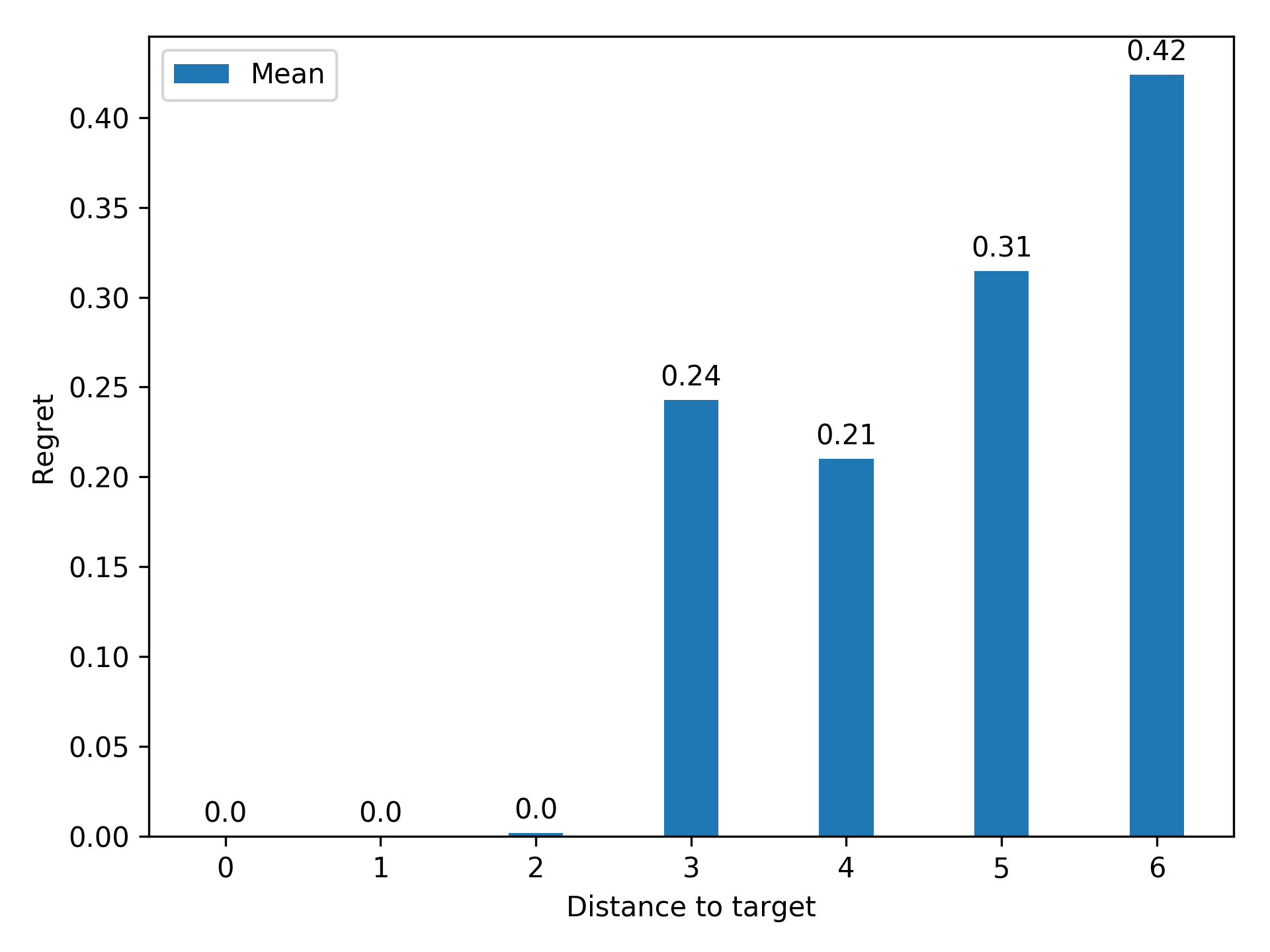}
  \caption{Dynamic regret analysis}
  \label{fig:regret}
\end{figure}

\begin{figure}[htbp]
  \centering
  \includegraphics[width=.9\linewidth]{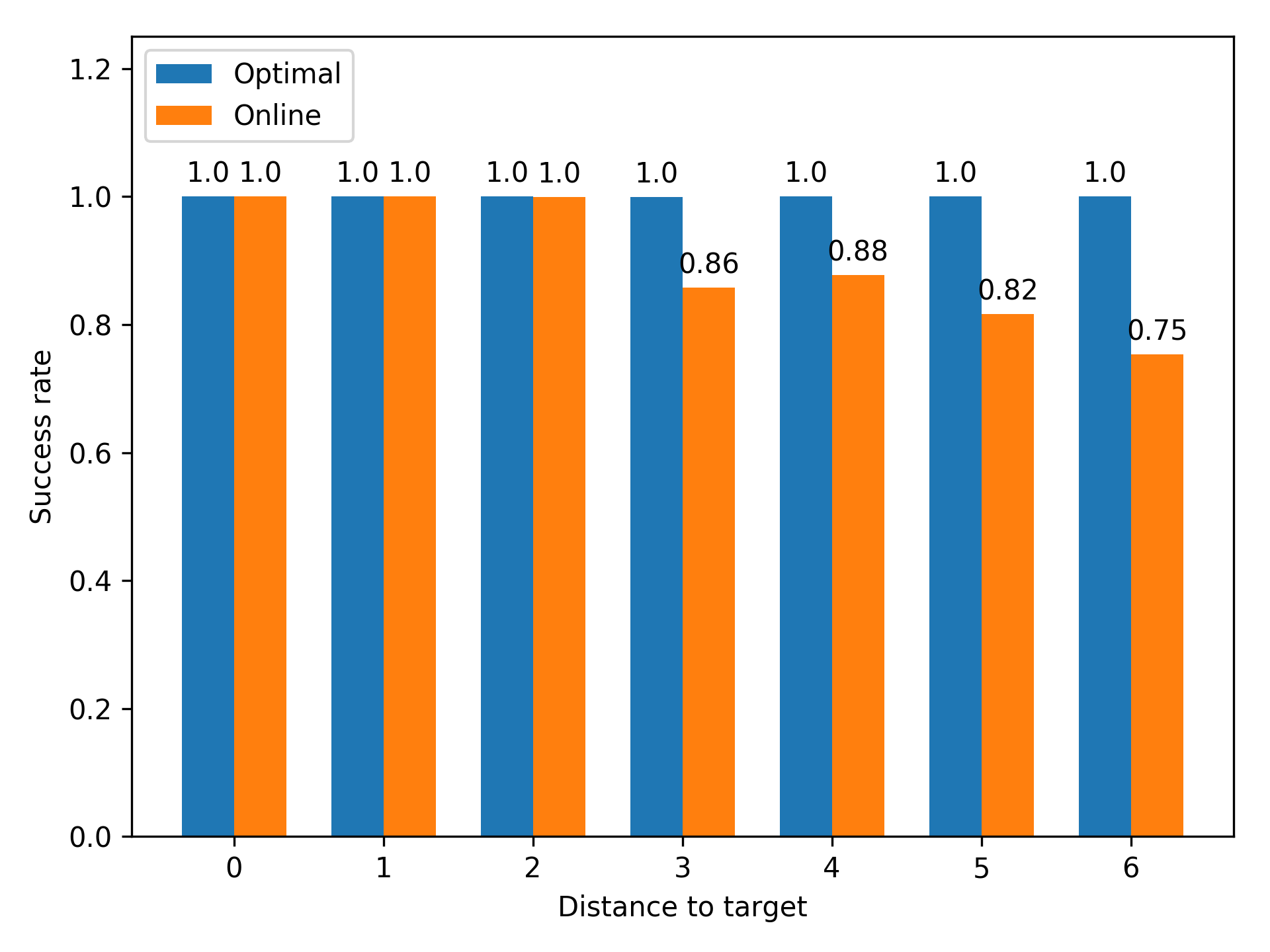}
  \caption{Success rate analysis}
  \label{fig:success rate}
\end{figure}

In this experiment, we further evaluate the effect of the distance of the attacker initial state to the target on the MTD policy. Optimal and online policies are computed for ten sequences of random IDS  configurations. In each IDS configuration, three IDSs are randomly sampled from the IDS set $\mathcal{N}$ at $\frac{1}{3}Hz$. Evaluation is performed for different initial positions of the attacker, i.e., $s_0\in \{7, 13, 9, 11, 10, 0, 19\}$ with the distances to the target (measured by the shortest path in the graph) ranging from zero to six, respectively. The final results are show in Fig. \ref{fig:regret} and \ref{fig:success rate}. Fig. \ref{fig:regret} shows the dynamic regrets computed according to \eqref{eq:regret-formula} with $h = 19$. Based on the mean value of the regrets, the closer the attacker's initial position is to the target, the smaller the regret is, and hence the less effective the MTD policy is against the attacker. Particularly, when the distance to target is smaller than three, the regret approaches zero and the MTD policy has almost no effect.  


Fig.~\ref{fig:success rate} compares the success rate for optimal and online policies at different attacker initial states. From Fig. \ref{fig:success rate}, it can be seen that with the optimal policy, regardless of the distance from the attacker initial node to the target, the attacker can always reach the target with a success rate of 1. On the other hand, with online policy, the attacker's success rate decreases as the distance to the target increases. Chi-squared test is performed on the two-way data set. The data are classified into two mutually exclusive classes: winning when the attacker reaches the target, and losing when the attacker fails to reach the target. The $p$ value is $8.2\times 10^{-13}$, indicating that there is indeed a strong correlation between the success rate and the distance to the target.

\section{Conclusions and Discussion}
\label{sec:conclude}
In this paper, we have introduced a method to evaluate the effectiveness of a \ac{mtd} policy to detect the presence of adversaries. Given time varying locations of detection systems in a network, we formulate planning problem for a stealthy attacker using receding horizon framework. The attacker repeatedly  performs reconnaissance to figure out where \ac{ids}s are placed and  solves a risk-sensitive finite-horizon planning problems based on a probabilistic attack graph. We have assessed the effectiveness of the proactive defense strategy using the detection rate in the presence of such an intelligent attacker. This work provides foundations for several future extensions. First, we will investigate an adaptive attacker, who learns the dynamics of the network from past iterations. Several no-regret learning algorithms and online planning in \ac{mdp}s with regret bounds can be considered  for attacker behavior modeling. Second, given the evaluation  result, we can construct the game between a defender, who selects subsets of nodes for randomization, and an intelligent, potentially adaptive attacker. Through game-theoretic reasoning, we can compute optimal detection strategy that trades off multiple objectives, including maximizing the detection rate and minimizing the operational cost.

\section*{Acknowledgment}
This material is based upon work supported by the Defense Advanced Research Projects Agency (DARPA) under Agreement No. HR00111990015.
This work is also partially supported by grants CNS-1544782, SES-1541164 and ECCS-1847056 from National Science Foundation (NSF), and by award 2015-ST-061-CIRC01, U. S. Department of Homeland Security.

\bibliographystyle{./bibliography/IEEEtran}
\bibliography{./bibliography/IEEEabrv,./bibliography/ztrefs,jfrefs}
\end{document}